\documentclass[prl,twocolumn,
showpacs,preprintnumbers,superscriptaddress,amsmath,amssymb,floatfix,aps]{revtex4}

\usepackage[dvips]{graphicx}
\usepackage{dcolumn}
\usepackage{bm}
\begin{document}

\title{Dark solitons in a superfluid Fermi gas}

\author{Mauro Antezza}
\affiliation{Dipartimento di Fisica, Universit\`a di Trento and
CNR-INFM BEC, Via Sommarive 14, I-38050 Povo, Italy}
\author{Franco Dalfovo}
\affiliation{Dipartimento di Fisica, Universit\`a di Trento and
CNR-INFM BEC, Via Sommarive 14, I-38050 Povo, Italy}
\author{Lev P. Pitaevskii}
\affiliation{Dipartimento di Fisica, Universit\`a di Trento and
CNR-INFM BEC, Via Sommarive 14, I-38050 Povo, Italy}
\affiliation{Kapitza Institute for Physical Problems, ul. Kosygina
2, 119334 Moscow, Russia}
\author{Sandro Stringari}
\affiliation{Dipartimento di Fisica, Universit\`a di Trento and
CNR-INFM BEC, Via Sommarive 14, I-38050 Povo, Italy}

\date{\today}

\begin{abstract}
We investigate the behavior of dark solitons in a superfluid Fermi gas along the BCS-BEC crossover by solving the Bogoliubov - de Gennes
equations and looking for real and odd solutions for the order
parameter. We show that in the resonance unitary region, where the
scattering length is large, the density profile of the soliton has a
deep minimum, differently from what happens in the BCS regime. The
superfluid gap is found to be significantly quenched by the presence
of the soliton due to the occurrence of Andreev fermionic bound
states localized near the nodal plane of the order parameter.
\end{abstract}
\pacs{03.75.Lm, 03.75.Ss, 03.75.Kk}

\maketitle

\section{Introduction}
The interplay between effects of coherence and interaction is one
of the most interesting features exhibited by superfluids. It
shows up, in particular, through the occurrence of topological
excitations such as vortices and solitons. These have been the
object of systematic investigations in the case of ultracold Bose
gases, where coherence is the result of Bose-Einstein condensation
(BEC) and is associated with long range order in the one-body
density matrix. The situation is even more interesting in the case
of Fermi gases where coherence originates from the interaction
between particles which, at low temperature, brings the system
into a superfluid phase characterized by long range order in the
two-body density matrix. For this reason, the consequences of
coherence on measurable quantities, like the density distribution
of the gas, are more indirect and subtle than for bosons.

Fermi superfluid gases are now available experimentally in
$^{40}$K and $^6$Li and various regimes along the BCS-BEC
crossover can be explored by tuning the atomic $s$-wave scattering 
length $a$ through the Feshbach resonances exhibited by these atoms. 
When $a$ is small and negative the system is described by the BCS
theory of superfluidity. Conversely when $a$ is small and positive
dimers of atoms of different spin are formed and the system
behaves like a BEC of molecules. Near resonance the scattering
length is much larger than the average interparticle distance and
the system enters the so called unitary regime. Quantized vortices
have been recently observed along the BCS-BEC crossover \cite{Zwierlein} 
and have been the object of several theoretical papers
\cite{Feder,Bulgac,Machida,RanderiaVortex,Chien}. 

In this work we investigate another important class of nonlinear
topological excitations: dark solitons. In three dimensions (3D) a
dark soliton is characterized by a real order parameter which changes
sign at a planar node (a point node in 1D).  In the BEC regime the
soliton is a solution of the Gross-Pitaevskii (GP) equation for
the order parameter of the condensate with repulsive interaction
\cite{Tsuzuki}. In a uniform 3D system dark solitons are known to
be unstable via the snake instability, i.e., a sinusoidal
transverse oscillation of the planar node. In trapped gases,
however, the instability timescale can be very long, so
that solitons can indeed be observed \cite{exptBEC}.

While in the BEC case the node of the order parameter causes a 
notch in the density distribution, in a BCS superfluid the density 
is almost unaffected by the presence of the node \cite{Dziarmaga}. 
The situation is similar to the one of vortices. As in that case, 
the behavior of the density along the BCS-BEC crossover is expected 
to be interesting and rather nontrivial, as a result of the delicate 
balance of coherence and nonlinear interactions. A major question 
concerns the behavior at unitary where no exact many-body theory is 
presently available. The problem also shares useful analogies with 
the interference between two expanding Fermi superfluid gases where 
the order parameter is expected to exhibit an oscillating behavior 
with a change of sign, but no quantitative predictions are available 
concerning the behavior of the density.

\section{Bogoliubov - de Gennes theory}
We investigate the problem by using a mean-field theory for a 3D
Fermi gas at zero temperature, based on the solution of the
Bogoliubov - de Gennes (BdG) equations \cite{BdG,LeggRand}:
\begin{equation}
\label{BdG}
\left[
\begin{array}{cc}
\hat{H} & \Delta({\bf r})\\
\Delta^*({\bf r}) & -\hat{H}
\end{array}
\right]\;
\left[
\begin{array}{c}
u_\eta({\bf r}) \\
v_\eta({\bf r})
\end{array}
\right]= \varepsilon_\eta\; \left[
\begin{array}{c}
u_\eta({\bf r}) \\
v_\eta({\bf r})
\end{array}
\right],
\end{equation}
where $\hat{H}=-\hbar^2\nabla^2/2M+V_{\rm ext}({\bf r})-\mu$  is
the single-particle grand-canonical Hamiltonian. For given
chemical potential, $\mu$, and order parameter of the superfluid 
phase, $\Delta({\bf r})$, these equations provide the  spectrum 
$\varepsilon_\eta$ and quasiparticle amplitudes $u_\eta({\bf r})$ 
and $v_\eta({\bf r})$, which are required to satisfy the normalization 
relation $\int \textrm{d}^3{\bf r}\left[u^*_{\eta'}({\bf r})
u_\eta({\bf r})+v^*_{\eta'}({\bf r})v_\eta({\bf r})\right]=
\delta_{{\eta'} \eta}$. The above equations must be solved together 
with the equations for the order parameter and the density:
\begin{eqnarray}
\Delta({\bf r}) & = & -g\sum_\eta u_\eta({\bf r})v_\eta^*({\bf r}),
\label{Delta} \\
n({\bf r}) & = & 2\sum_\eta |v_\eta({\bf r})|^2 .
\label{density}
\end{eqnarray}
This is done by means of an iterative procedure, which starts from
a trial function $\Delta({\bf r})$ and converges to the
self-consistent solution of (\ref{BdG})-(\ref{density}). All sums
in the equations are limited by an energy cutoff,
$0 \leq \varepsilon_\eta \leq E_c$. This cutoff is required
in order to cure the ultraviolet divergences and is accompanied by
a regularization of the interaction parameter $g$ according to
\cite{LeggRand}:
\begin{equation}
\frac{1}{k_Fa}= \frac{8\pi
\varepsilon_F}{gk_F^3}+\frac{2}{\pi}\sqrt{\frac{E_c}{\varepsilon_F}}.
\label{eqg}
\end{equation}
Here $a$ is the 3D $s$-wave scattering length
characterizing the interaction between atoms of different spins,
while $\varepsilon_F=\hbar^2k_F^2/2M$  and $k_F=(3\pi^2n_0)^{1/3}$ 
are the Fermi energy and momentum of a uniform ideal Fermi gas of
density $n_0$ \cite{note-kf}. For sufficiently large values of 
$E_c$ the final results should not depend on the choice of the 
cutoff. It is worth stressing that the above equations reduce to
the stationary GP equation for the order parameter of a condensate
of molecules in the BEC limit \cite{Pieri}, the interaction
between dimers being however given by the mean field value $2a$
instead of the exact value $0.6 a$ \cite{Petrov}. Although
approximate, this mean field theory is expected to give a
comprehensive and reasonably accurate picture of the BCS-BEC
crossover.

\section{Results for a Dark Soliton}
We look for solutions of (\ref{BdG})-(\ref{density})
corresponding to a soliton at rest in the superfluid and in
the absence of external potential ($V_{\rm ext}=0$)
\cite{note-nonlinear}. The order parameter, $\Delta(z)$, is chosen
to be a real and odd function of $z$, having a node in the
$xy$-plane at $z=0$. The density $n(z)$ is an even
function of $z$. The condition of reality implies the absence of
currents, which are in general associated with a $z$-dependence of
the phase of the order parameter. The calculation is done in a
finite box of size $L \times L_{\perp}^2$, where $L$ is the size
along $z$ and all solutions are forced to vanish at the boundaries. 
If the box size is large enough the effects of the boundaries on 
the soliton are vanishingly small. Due to the translational symmetry 
in the transverse direction, the BdG eigenfunctions can be written 
as $u_\eta({\bf r})= u_{n,n_\perp}(z)\ e^{i {\bf k}_\perp \cdot {\bf
r}_\perp}\; \sqrt{k_F}/L_{\perp}$ and $v_\eta({\bf r})=
v_{n,n_\perp}(z)\ e^{i {\bf k}_\perp \cdot {\bf
r}_\perp}\;\sqrt{k_F}/L_{\perp}$, where ${\bf r}_\perp = (x,y)$
and ${\bf k}_\perp = (k_x,k_y)$. The transverse momentum is
quantized according to the rule $k_x=2\pi n_x/L_{\perp}$ and
$k_y=2\pi n_y/L_{\perp}$, with $n_\perp = (n_x,n_y)$ and $n_x$,
$n_y$ integers. We consider a gas in a box with $L=40k_F^{-1}$ and
$L_{\perp}=20k_F^{-1}$, and we also set $E_c=50\varepsilon_F$, 
which turns out to be large enough for a reasonable convergence.

\begin{figure}[ptb]
\begin{center}
\includegraphics[width=0.47\textwidth]{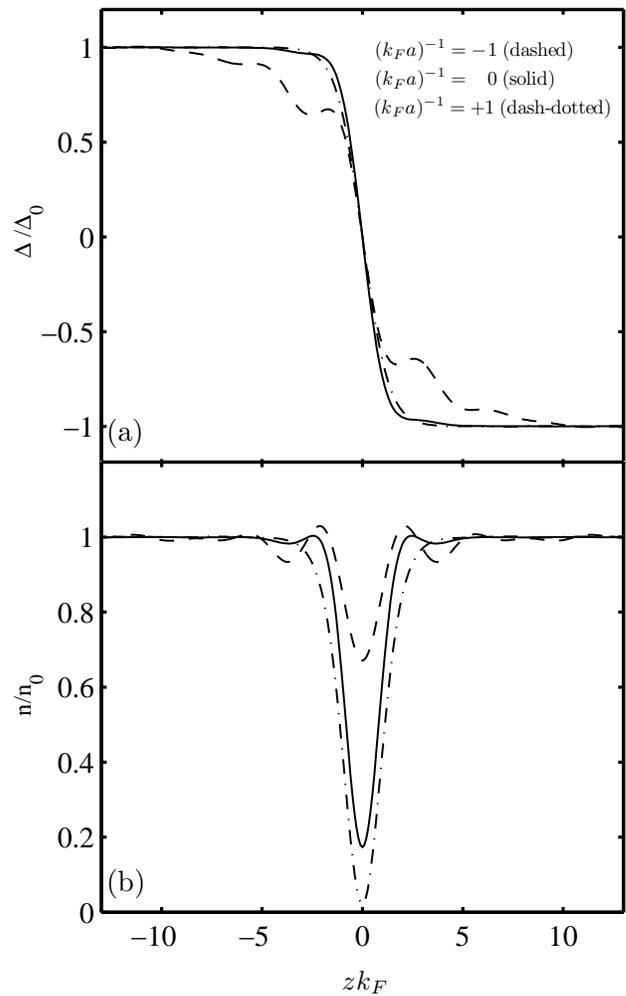}
\caption{\footnotesize{$(\textrm{a})$ Order parameter and $(\textrm{b})$ density
for a dark soliton with $(k_Fa)^{-1}=0$ and $\pm 1$. The value
$(k_Fa)^{-1}=0$ (solid line) corresponds to unitarity, while
$(k_Fa)^{-1}=-1$ (dashed) and $(k_Fa)^{-1}=+1$ (dot-dashed) are on
the BCS and BEC side of the resonance, respectively. Both
$\Delta(z)$ and $n(z)$ are normalized to their asymptotic values
far away from the soliton. }}
\label{Fig:gapdens}
\end{center}
\end{figure}

In Fig.~\ref{Fig:gapdens} we show the order parameter (a) and the
density (b) for a dark soliton with $(k_Fa)^{-1}=-1$, $0$ and $+1$.
Both quantities are normalized to their asymptotic values far away
from the soliton, $n_{0}= k_F^3/3\pi^2$ and $\Delta_{0}$. For 
$\Delta_{0}$ and $\mu$ we find the values: $\Delta_{0}\simeq
0.21\varepsilon_F$ and $\mu\simeq 0.96\varepsilon_F$ for
$(k_Fa)^{-1}=-1$ (BCS side); $\Delta_{0}\simeq 0.70\varepsilon_F$
and $\mu\simeq 0.61\varepsilon_F$ for $(k_Fa)^{-1}=0$
(unitarity); $\Delta_{0}\simeq 1.5\varepsilon_F$ and
$\mu\simeq -0.84\varepsilon_F$ for $(k_Fa)^{-1}=+1$ (BEC side).
These values almost coincide with those of an infinite uniform 
system when calculated with the same value of the cutoff energy. 
This proves that the size of the box is large enough to neglect 
its effects on the calculations. They instead differ from the 
values calculated in a uniform gas in the limit of infinitely 
large $E_c$. The difference is about $\simeq 10\%$ on the BEC 
side and $\simeq 1\%$ in the other cases. A larger discrepancy 
is of course obtained by using a lower cutoff energy, especially 
on the BEC side of the resonance where the formation of molecules 
with energy $\sim \hbar^2/ma^2$ requires large values of $E_c$ 
to reach convergence.

Figure \ref{Fig:gapdens}b shows the occurrence of a deep depletion 
of the density at unitarity (solid line) with a $\simeq 80\%$
contrast, comparable to the one of the BEC regime (dash-dotted). 
On the BCS side (dashed), conversely, the contrast is only 
$\simeq 30\%$ at $(k_Fa)^{-1}=-1$ and becomes exponentially small 
in the limit $k_F|a| \ll 1$.  These results are consistent with 
those obtained for the profile of a vortex 
core \cite{RanderiaVortex,Chien}.

In Fig.~\ref{Fig:gapdens}a one notices that, as for the core of a
vortex, the order parameter of the soliton in the BCS regime
exhibits two length scales: a steep slope in a narrow region of
the order of $k_F^{-1}$ and a smoother slope in wider region of
size $\xi_{\rm BCS}=\hbar v_F/\Delta_0$, where $\xi_{\rm BCS}$ is
the coherence length of the Cooper pairs and $v_F=
\sqrt{2\varepsilon_F/M}$ is the Fermi velocity. For $(k_Fa)^{-1}
=-1$ one has $\xi_{\rm BCS} \simeq 10 k_F^{-1}$. We also find 
that the density exhibits oscillations with wavevector $\sim 2 k_F$. 
The same type of oscillations are found at the box boundaries and we
checked that their shape does not depend on neither the box size
nor the cutoff energy. One can thus safely identify them as
Friedel oscillations. Approaching unitarity the coherence length
$\xi_{\rm BCS}$ decreases, eventually reducing to $k_F^{-1}$. 

\begin{figure}[ptb]
\begin{center}
\includegraphics[width=0.48\textwidth]{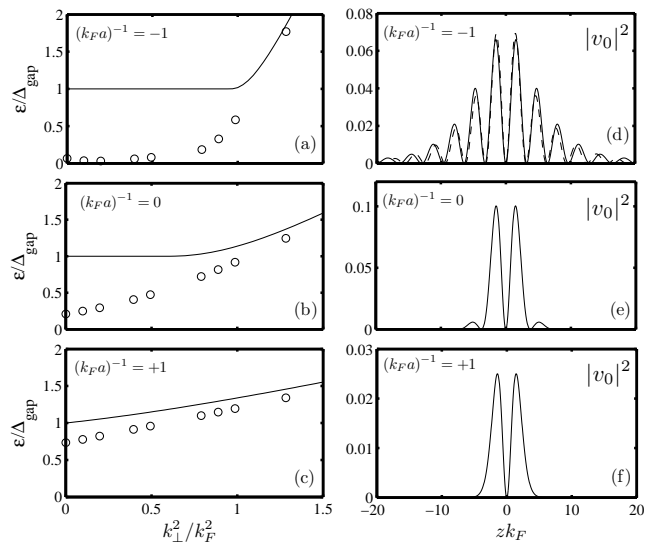}
\caption{\footnotesize{
Left panels: Energy of the lowest single-particle states for fixed
$k_{\perp}$ and for (a) 
$(k_Fa)^{-1}=-1$, (b) $0$ and (c) $+1$. Circles correspond to the lowest
bound states in presence of the dark soliton; solid lines correspond to the lowest
energy states, $\varepsilon_{\textrm{bulk}}$, in a uniform superfluid (see text). Right panels: 
$|v_{0}(z)|^2$ for the lowest bound state with $k_{\perp}=0$.
The dashed line in (d) corresponds to the ansatz
$v_{0}(z) \sim \sin(k_F z) \exp (z/\xi_{\rm BCS})$
with $\xi_{\rm BCS} = 10 k_F^{-1}$.}}
\label{Fig:spectra}
\end{center}
\end{figure}

Let us focus now on the single-particle excitation spectrum.
In the uniform superfluid the spectrum is given by the well known 
result $\varepsilon_{\textrm{bulk}}(k) = [(\hbar^2k^2/2M
-\mu)^2+\Delta_0^2]^{1/2}$ \cite{nobosonic}. This expression is
plotted with solid lines in Fig.~\ref{Fig:spectra}(a-c) as a 
function of $k^2_\perp$, for the three cases $(k_Fa)^{-1}=0$ and 
$\pm 1$. The minimum of this function defines the gap
$\Delta_{\textrm{gap}}$ for bulk excitations, which is
$\Delta_{\textrm{gap}}=\Delta_0$ for $\mu>0$ and
$\Delta_{\textrm{gap}}=[\mu^2+\Delta_0^2]^{1/2}$ for $\mu<0$. 
In the presence of the soliton, however, the eigenvalues
$\varepsilon_{n,n_\perp}$ of the BdG equations exhibit a
nontrivial feature: besides the continuum of bulk states with 
energy above $\varepsilon_{\textrm{bulk}}$ one finds several 
states even below $\varepsilon_{\textrm{bulk}}$.  The energy of
the lowest states for each $k_\perp$ is given in 
Fig. ~\ref{Fig:spectra}(a-c) (open circles). The corresponding 
quasiparticle amplitudes are localized near the soliton, as one 
can see in Fig.~\ref{Fig:spectra}(d-f) where we plot the 
function $|v_{0}(z)|^2$ of the state with $k_\perp=0$. The origin 
of these localized Andreev-like states \cite{AndreevRefl} 
resides in the fact that the energy cost for creating a fermionic 
excitation near a node of the order parameter is reduced with 
respect to the bulk value. In the BCS limit the energy of the 
lowest bound state, also called {\it minigap} in the context 
of superconductivity, is expected to be of the order of
$\Delta_0^2/2\varepsilon_F$. For a vortex in a BCS superfluid 
described by BdG equations this result was proved in 
Ref.~\cite{Caroli}, but the result is rather general and can 
be derived also when $\Delta(z)$ is a step function 
\cite{Yoshida,Mueller}. In the same limit, the eigenvectors 
$u_0(z)$ and $v_0(z)$ of the lowest bound state behave like 
$\sim \cos(k_F z) \exp (z/\xi_{\rm BCS})$ and $\sim \sin(k_F z) 
\exp (z/\xi_{\rm BCS})$, respectively, as
shown in Fig.~\ref{Fig:spectra}d. For the minigap at $(k_Fa)^{-1}=-1$
we find  $\Delta_{\rm minigap} \simeq 0.64 \Delta_0^2/2\varepsilon_F$.

\begin{figure}[ptb]
\begin{center}
\includegraphics[width=0.41\textwidth]{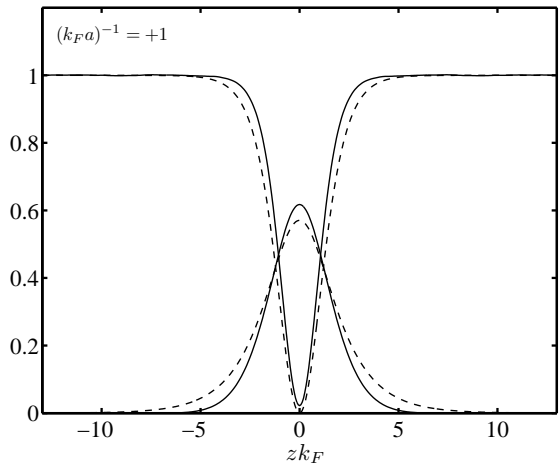}
\caption{\footnotesize{Dashed lines: density profile $n_B(z)/n_{B0}$ and wave 
function of the lowest fermionic bound state, $\Psi_0(z)$, in the 
BEC limit. Solid lines: density profile $n(z)/n_0$ and lowest 
eigenvector $u_0(z)$ of the BdG equations for $(k_Fa)^{-1}=+1$. }}
\label{Fig:andreev}
\end{center}
\end{figure}

We now discuss the BEC limit, where the density profile 
should approach the analytic result obtained by solving the GP
equation for a gas of bosons with mass $m_B=2M$ interacting with 
a scattering length $a_{BB}$. The density of the bosonic 
dimers, $n_B=n/2$, is given by \cite{Tsuzuki}:  
$n_B(z)= n_{B0} \tanh^2(z/\sqrt{2}\xi_{\rm BEC})$,  
with $n_{B0} = n_0/2=k^3_F/(6\pi^2)$, $\xi_{\rm BEC}=\hbar/
\sqrt{2m_Bg_{BB}n_{B0}}$ and $g_{BB}=4\pi \hbar^2a_{BB}/m_B$.
Analytic solutions can be found, in 
the same limit, also for the fermionic bound states. In fact, 
these states can be obtained by solving a Schr\"odinger equation 
for a fermionic impurity of mass $m_F=M$ in an inhomogeneous 
bosonic superfluid. By minimizing the total energy functional 
one derives the following equation for the impurity wave 
function $\Psi$: 
\begin{equation}
\left[-\frac{\hbar^2}{2m_F} \partial^2_z + g_{BF}n_B(z)\right]
\Psi(z)=\varepsilon \Psi(z),
\label{Andreev}
\end{equation}
where $g_{BF}=2\pi\hbar^2a_{BF}/m_r$ is the dimer-atom  
coupling constants, while $m_r=m_Bm_F/(m_B+m_F)=2M/3$ is the 
reduced mass. The exact values of the $a_{BB}$ and $a_{BF}$ 
scattering lengths are equal to $0.6 a$ \cite{Petrov} and 
$1.2 a$ \cite{Skorniakov}, respectively, while the BCS mean 
field theory yields the Born approximation values
$2a$ and  $8/3 a$ \cite{Pieri2}. Equation~(\ref{Andreev})
corresponds to the Schr\"odinger equation for a particle in a
modified P\"oschl-Teller potential. The solutions are analytic
\cite{Fluegge} and include bound states. The energy of these bound
states can be used to calculate the minigap for fermionic
excitations in the BEC regime. In fact the breaking of a pair
causes the unbound motion of two fermions which will occupy the
single particle states $\Psi_0$ with lowest energy. Using the mean
field values for the scattering length, the energy of the lowest
state turns out to be exactly one half of the bulk value
$g_{BF}n_0/2$ \cite{note-exact} so that
\begin{equation}
\Delta_{\rm gap}-\Delta_{\rm minigap} = 4 \pi \hbar^2 a n_0/M \, .
\label{gain}
\end{equation}
Equation (\ref{gain}) provides a good quantitative estimate for
the gap even for $(k_Fa)^{-1}=+1$ where the above expression gives the
value $\Delta_{\rm gap}- \Delta_{\rm minigap}= 0.26 \Delta_{\rm
gap}$ to be compared with our numerical result $\simeq 0.27 
\Delta_{\rm gap}$. In Fig.~\ref{Fig:andreev} the analytic curve 
for the density $n_B(z)$ and the bound state wave function $\Psi_0(z)$ 
(dashed lines) are compared with $n(z)/2$ and $u_0(z)$ obtained 
from BdG equations (solid lines). The agreement is rather good and 
the small differences are due to the fact that $(k_Fa)^{-1}=+1$ is 
still relatively far from the true BEC limit and to the use of a finite 
cutoff energy $E_c$.

\section{Conclusions}
In conclusion, we have theoretically investigated the behavior of
dark solitons in a Fermi superfluid in the BCS-BEC crossover,
showing that the soliton at unitarity has a large density
contrast. We have also discussed the existence of bound
states. Fermionic bound states have already been the object of 
investigations in the context of 1D dark solitons in conducting 
polymers \cite{Heeger}. They play an important role also in the 
physics of domain walls in the Fulde-Ferrell-Larkin-Ovchinnikov 
states (see, for instance, Ref.~\cite{Ichioka,Yoshida} and references 
therein) and in the case of Josephson currents through a potential
barrier \cite{Spuntarelli}. In the opposite BEC limit, the binding 
of the unpaired Fermi atoms on the solitonic plane shares an 
interesting analogy with the Andreev state of $^3$He atoms on the
free surface of superfluid $^4$He \cite{AndreevSufTen}. In the 
geometry of the soliton, the fermionic bound state can give rise to 
a 2D Fermi gas embedded into a molecular Bose superfluid. This can be
particularly interesting when the molecular BEC is made starting
from a slightly imbalanced spin population. The residual gas of
unpaired atoms can easily fill the available states in the
soliton, thus forming a polarized 2D Fermi gas.
From the experimental viewpoint, it seems quite possible
to observe solitons in the BCS-BEC crossover, for instance, by 
producing them in the molecular BEC phase with known techniques 
\cite{exptBEC} and then tuning the scattering length across the 
Feshbach resonance, similarly to what has already been done with 
vortices. A natural extension
of this work is the study of solitons characterized by a complex order parameter and moving
in the superfluid (grey solitons).

\section{Acknowledgments}
We thank S. Giorgini, E.J Mueller, and M. Randeria for 
stimulating discussions. We acknowledge supports by 
the Ministero dell'Universit\`a e della Ricerca (MiUR).



\begin{thebibliography}{99}

\bibitem{Zwierlein}
M. W. Zwierlein  {\it et al.},
Nature (London) {\bf 435}, 1047 (2005).

\bibitem{Feder}
N. Nygaard {\it et al.},
\prl {\bf 90}, 210402 (2003).

\bibitem{Bulgac}
A. Bulgac, Y. Yu,
\prl {\bf 91}, 190404 (2003).

\bibitem{Machida}
M. Machida and T. Koyama,
\prl {\bf 94}, 140401 (2005).

\bibitem{RanderiaVortex}
R. Sensarma, M. Randeria, T.-L. Ho,
\prl {\bf 96}, 090403 (2006).

\bibitem{Chien}
C.-C. Chien {\it et al.},
Phys. Rev. A {\bf 73}, 041603(R) (2006).

\bibitem{Tsuzuki}
T. Tsuzuki,
J. Low Temp. Phys. {\bf 4}, 441 (1971).

\bibitem{exptBEC}
S. Burger {\it et al.},
\prl {\bf 83}, 5198 (1999);
J. Denschlag {\it et al.},
Science {\bf 287}, 97 (2000);
B.P. Anderson {\it et al.},
\prl {\bf 86}, 2926 (2001);
Z. Dutton {\it et al.},
Science {\bf 293}, 663 (2001);
N.S. Ginsberg, J. Brand, and L.V. Hau,
\prl {\bf 94}, 040403 (2005);
P. Engels and C. Atherton,
e-print~arXiv:0704.2427;
J. Brand, L.D. Carr, B.P. Anderson,
e-print~arXiv:0705.1341 (2007).

\bibitem{Dziarmaga}
J. Dziarmaga and K. Sacha,
Laser Phys. {\bf 15} (4), 674 (2005).

\bibitem{BdG}
P.G. de Gennes,
{\it Superconductivity of Metals and Alloys},
(Benjamin, new York, 1966).

\bibitem{LeggRand}
A.J. Leggett
in {\it Modern Trends in the Theory of Condensed Matter},
edited by A. Pekalski and R. Przystawa
(Springer-Verlag, Berlin, 1980);
M. Randeria
in {\it Bose Einstein Condensation},
edited by A. Griffin, D. Snoke, and S. Stringari
(Cambridge, Cambridge, England, 1995).

\bibitem{note-kf} In typical experiments with 
ultracold gases $n_0 \sim 10^{11}\div 10^{12}$ 
atoms/cm$^3$ corresponding to $k_F^{-1}$ of 
few $\mu$m.

\bibitem{Pieri}
P. Pieri and G. Strinati,
\prl {\bf 91}, 030401 (2003).

\bibitem{Petrov}
D.S. Petrov, C. Salomon, G.V. Shlyapnikov,
\prl {\bf 93}, 090404 (2004).

\bibitem{note-nonlinear} The existence of nonuniform solutions, such
as solitons and vortices, even in the absence of external potential
is a consequence of the nonlinear character of the self-consistent
equations (\protect\ref{BdG})-(\protect\ref{density}).


\bibitem{nobosonic} We remind here that BdG equations (\ref{BdG})-(\ref{density}) account for
fermionic single-particle excitations and do not include bosonic
collective excitations.

\bibitem{AndreevRefl} 
A.F. Andreev, Zh. Eksp. Teor. Fiz. {\bf 46}, 1823 (1964) [Sov. Phys. 
JETP {\bf 19}, 1228 (1964)]; \emph{ibid.} {\bf 49}, 655 (1965) 
[{\bf 22}, 455 (1966)].

\bibitem{Caroli}
C. Caroli, P. de Gennes, and J. Matricon,
Phys. Lett. {\bf 9}, 307 (1964)


\bibitem{Ichioka}
T. Mizushima, K. Machida, and M. Ichioka, \prl
{\bf 94}, 060404 (2005).

\bibitem{Yoshida}  
N. Yoshida and S.-K. Yip, \pra {\bf 75}, 063601 (2007).

\bibitem{Mueller}
E.J. Mueller, private communication.

\bibitem{Skorniakov} G.V. Skorniakov and K.A. Ter-Martirosian,
Sov. Phys. JETP {\bf 4}, 648 (1957).

\bibitem{Pieri2} P. Pieri and G.C. Strinati, \prl
{\bf 96}, 150404 (2006).

\bibitem{Fluegge} S. Fl\"ugge, {\it Practical
Quantum Mechanics} (Springer-Verlag, Berlin, 1994).

\bibitem{note-exact} By using the exact values $a_{BB}=0.6 a$
and $a_{BF}=1.2 a$ the ratio between the lowest bound state
and bulk state energies is $0.43$ instead of $0.5$, and a
second weakly bound state appears with a ratio $0.97$.

\bibitem{Heeger} A. J. Heeger {\it et al}, Rev. Mod. Phys. 
{\bf 60}, 781 (1988).

\bibitem{Spuntarelli}
A. Spuntarelli, P. Pieri, G.C. Strinati, \prl {\bf 99}, 040401 (2007).

\bibitem{AndreevSufTen} 
A. F. Andreev, Zh. Eksp. Teor. Fiz. {\bf 50}, 1415 (1966); 
[Sov. Phys. JETP {\bf 23}, 939 (1966)]. 


\end{thebibliography}
\end{document}